\documentstyle[prl,graphicx,aps]{revtex}

\newcommand{\cross}  {\times}
\newcommand{\diff}   {{\rm\, d}}

\newcommand{\beq}    {\begin{equation}}
\newcommand{\enq}    {\end{equation}}

\newcommand{\pt}     {\varphi}

\begin{document}

\title{Collective transport properties of bilayer-quantum-Hall excitonic condensates}
\author{Allan H. MacDonald, Anton A. Burkov, Yogesh N. Joglekar, and Enrico Rossi} 
\address{Department of Physics, The University of Texas at Austin, Austin, TX 78712, USA}
\date{\today}
\maketitle

\begin{abstract}
Double-layer electron systems in the quantum Hall regime have excitonic condensate ground states 
when the layers are close together and the total Landau level filling factor is close to an odd 
integer. In this paper we discuss the microscopic physics of these states and recent progress 
toward building a theory of the anomalous collective transport effects that occur in separately 
contacted bilayers when the condensate is present. 
\end{abstract}


\section{Introduction}
Among the broken symmetry states that occur in many-particle systems, those in which long range 
phase coherence is established, either for bosons\cite{becref} or for pairs of 
fermions\cite{bcstheory}, have a special significance because of the quantum nature of their
macroscopic order and the sometimes startling phenomenology that results. In semiconductors, the 
possibility of long-range phase coherence due to Bose condensation of excitons was first 
raised\cite{keldysh} nearly 40 years ago. In this paper, we argue that the anomalous transport 
properties discovered in bilayer quantum Hall systems by J.P. Eisenstein and 
collaborators\cite{jpeexpt} and studied in this group over the past couple of years, are an 
experimental manifestation of such an ordered state. We also briefly discuss the microscopic 
physics that causes this order to occur in the quantum Hall regime and efforts, currently in 
progress, to achieve a more complete understanding\cite{firstheory} of the collective transport 
effects that lie behind the experimental observations. This paper focuses on work related to 
these questions performed at the University of Texas. 


\section{Spontaneous Phase Coherence {\em is} Excitonic Bose Condensation} 
Excitonic Bose condensation in semiconductors has usually been discussed by describing the 
fermionic degrees of freedom of a valence band in terms of missing electrons, holes, and the 
fermionic degrees of freedom of a conduction band directly in terms of electronic states. Under 
this particle-hole transformation a valence-band electron creation operator 
($c_{v,\vec k}^{\dagger}$) is mapped to a hole annihilation operator ($d_{v,-\vec k}$). Here we 
have, for simplicity, dropped the spin degree-of-freedom. The subscript $v$ is a band label that 
specifies the valence band; theoretical discussions of excitonic Bose condensation often focus on 
models with a single parabolic valence band. In electron-hole language, the phase coherence of 
electron-hole pairs in an excitonic condensate state proposed by Keldysh and Koslov\cite{keldysh} 
is signaled by a finite value for the expectation value 
$\langle c_{c,\vec k}^{\dagger} \; d_{v,-\vec k}^{\dagger} \rangle$.
This quantity is closely analogous to the finite expectation value in superconductors for the 
operator which creates a pair of electrons in time-reversed states.  

Two main difficulties have caused the experimental realization of this idea to be a challenge. 
Firstly, the ground state of a semiconductor has no electrons or holes present; it is necessary 
to generate the electron-hole plasma optically. This doesn't solve the problem completely 
however, since electrons and holes can also recombine optically; the electron-hole plasma that is 
created is always in a somewhat complex non-equilibrium state. Additionally, even if electrons 
and holes are present and optical recombination does not preempt their condensation, differences 
between conduction and valence band dispersions may be\cite{conti} an obstacle. The problems 
presented by electron-hole recombination can be mitigated to a large degree by placing electrons 
and holes in separate quantum wells.  This idea was first suggested some time ago\cite{ehbilayer},
and recent experiments on optically generated electron-hole plasmas in such systems have uncovered 
some interesting ununderstood properties\cite{ehbilayernature}. To date, however, there has not 
been any compelling evidence that a phase coherent condensate has been achieved in the absence 
of magnetic field. We will argue below that over the past couple of years excitonic pair 
condensation has however been achieved in the presence of an external magnetic field, although it 
has not generally been recognized as such. 

The lack of recognition is, in our view, mainly a confusion of language. The situation is most 
clear if we consistently stick with most physically transparent description, instead of making a 
particle-hole transformation for the valence band, in describing the ordered state. The order 
parameter of the excitonic condensate state is then 
$\langle c_{c,\vec k}^{\dagger}\; c_{v,\vec k} \rangle$. Excitonic condensation is nothing but 
the development of spontaneous phase coherence between two different bands in a solid, spontaneous 
coherence in the case of bilayers between electron bands localized in two different quantum 
wells. This is precisely what occurs for bilayer electron systems in the quantum Hall regime near 
odd integer total filling factors. The prediction that this broken symmetry would occur in 
bilayer quantum Hall systems\cite{fertig,macd,jungw}, early experimental 
indications\cite{macd,murphy} of coherence-dependent Hall gaps, and many aspects of the theory of 
these states\cite{bilayertheory} were, however, developed largely\cite{elhole} without reference 
to the literature on excitonic Bose condensation. Because their spontaneous coherence occurs 
between two different two-dimensional {\em conduction subbands}, there is no need to 
simultaneously populate the conduction band with electrons and the valence band with holes in 
order to establish conditions under which the broken symmetry can occur. This eliminates one key 
difficulty that has plagued efforts to realize excitonic condensation. At the same time 
dispersionless Landau bands eliminate the requirement of Fermi surface nesting\cite{conti}.


\section{Why Excitonic Condensation between Conduction Bands is likely only in the Quantum Hall 
Regime}
In this section, we use mean-field theory to explain why excitonic condensation (spontaneous 
coherence) at zero magnetic field is more likely between quasi-2D subbands that have opposite 
energy-wavevector dispersions, {\it i.e.} between conduction and valence subbands, and why this 
tendency no longer applies in the strong magnetic field limit. In mean-field theory, the ordered 
states we are interested in are Slater determinants formed from quasiparticles with spontaneous 
coherence and are closely analogous to the BCS variational wavefunctions familiar from the 
theory\cite{bcstheory} of superconductivity. Thus
\begin{equation}
\vert \Psi \rangle = \prod_{\vec k \in S} [u_{\vec k} \; c^{\dagger}_{T,\vec k}
+ v_{\vec k} \; c^{\dagger}_{B,\vec k}] \prod_{\vec k \in D} c^{\dagger}_{T,\vec k} \; 
c^{\dagger}_{B,\vec k} \vert 0 \rangle
\label{eq:vwfbzero}
\end{equation}
where $\vec k$ is a Bloch wavevector, $|u_{\vec k}|^2+|v_{\vec k}|^2=1$, $\vec k\in S$ and 
$\vec k\in D$ are exclusive sets of singly and doubly occupied wavevectors, and the phase 
relationship between $u_{\vec k}$ and $v_{\vec k}$ is fixed. Here the subscript $T$ labels 
{\em top} layer electrons while $B$ labels {\em bottom} layer electrons. This form of variational 
wavefunction is more general than the simple BCS variational wavefunction in that allows some 
wavevectors $\vec k \in D$ to be occupied by a pair of quasiparticles; these wavevectors do not 
contribute to spontaneous coherence. Unlike a Fermi gas, which has all wavevectors inside the Fermi
sea occupied in both layers, top and bottom layer electrons in this variational wavefunction are 
correlated and have a reduced probability of being close together. The fact that these 
correlations are present can be verified by an explicit calculation or understood in terms of the 
fact that two electrons cannot have the same wavevector $\vec k \in S$ whether they are in the 
same layer or opposite layers. This state has spontaneous phase coherence when the set 
$\vec k\in S$ is not empty. Spontaneous coherence is established in order to improve interlayer 
correlations, but comes at a cost in intralayer correlations and kinetic energy. 

The dependence of this competition on band dispersion can be understood qualitatively from the 
following considerations. The cost in kinetic energy tends to be minimized when there is a set of 
wavevectors for which single occupation occurs even in the non-interacting electron ground state. 
For example, for parabolic bands and equal densities of conduction band electrons and valence 
band holes, one 
state is occupied at each Bloch wavevector in the non-interacting ground state. When the two 
bands have the same dispersion, however, either the two bands have a large energetic offset which 
works against order or the kinetic energy is minimized by occupying Bloch states for both bands 
at most wavevectors. This is not to say that spontaneous coherence between Bloch bands in a solid 
that have the same curvature is impossible; in fact ferromagnetism can be regarded as being due 
to spontaneous coherence between bands that are distinguished only by their spin index. (This 
view is natural when the spin-quantization axis is chosen to be in the plane perpendicular to the 
spontaneous magnetization direction.) Whenever the conditions are ripe for this type of order, 
however, it will usually be favorable to to establish coherence between bands with the same 
orbital label rather than different ones; ferromagnetism preempts spontaneous coherence between 
two conduction bands. In the quantum well case we know that ferromagnetism occurs within a 
single layer only at unattainably low electron densities, if at all. Spontaneous coherence between 
two-dimensional conduction subbands localized in different layers is therefore extremely unlikely.
Spontaneous coherence between conduction and valence subbands in different layers is much more 
likely, perhaps even probable, but hard to realize in experiment. All this changes in the quantum 
Hall regime. 

We now turn our attention to a bilayer quantum Hall system with total filling factor $\nu=1$, the 
most favorable circumstance for spontaneous interlayer phase coherence. For simplicity we assume 
that the magnetic field is strong enough that all electrons in both layers are in the 
lowest kinetic energy, majority spin, Landau level. The mean-field variational wavefunction for a 
phase coherent bilayer state has the form:
\begin{equation}
\vert \Psi \rangle = \prod_{k} [u_k c^{\dagger}_{T,k}
+ v_k c^{\dagger}_{B,k}] \vert 0 \rangle
\label{eq:gswfa}
\end{equation}
where $|u_k|^2+|v_k|^2=1$ and $k$ is the one-component wavevector that labels states in the 
lowest Landau level in Landau gauge. In this case it turns out that, partly because of the 
degeneracy of the Landau bands, $u$ and $v$ are independent of $k$ in the translational invariant 
mean-field ground state. Since there is one occupied orbital for each $k$, this state indeed has 
Landau level filling factor $\nu=1$. The mean-field state for the case of a conduction band 
Landau level in one layer and a valence band Landau level in the other layer would be identical, 
with $u$ and $v$ independent of $k$ in the ground state and no Landau gauge orbitals either empty 
or doubly occupied. Since there is no dispersion of the Landau band in either case, there is no 
reason to expect any qualitative dependence on the microscopic band from which the Landau level
is derived. This important property was apparently not appreciated in the literature\cite{pru}
on excitonic condensation in strong fields, likely because it is exclusively couched in the 
language of conduction band electrons and valence band holes. (For $\nu\ne 1$, some states in the 
variational wavefunction must be either empty or doubly occupied, weakening the tendency toward 
order as discussed above. The physics of this suppression is, we believe, quite sensitive to the 
disorder present in the quantum wells.) When the two subband energies are identical, the 
mean-field state minimizes the kinetic energy for any choice of its variational parameters, $u_k$ 
and $v_k$, and also, by establishing interlayer phase coherence, reduces the interlayer 
interaction energy. It turns out\cite{dlreview} that this mean-field wavefunction with 
$u_k = v_k = \exp (i \phi)/\sqrt{2}$ and arbitrary $\phi$ approaches the exact ground state of 
$\nu = 1$ bilayer quantum Hall systems for layer spacing $d \to 0$. Corrections to 
mean-field-theory can\cite{joglekarfluc} be calculated perturbatively for small $d$, but for 
large $d$, quantum fluctuation corrections to the mean-field state become substantial and 
eventually a phase transition occurs to a state without spontaneous coherence.


\section{Collective Transport Properties}
Bilayer quantum Hall ferromagnets have most often been described as pseudospin ferromagnets. If 
states in the top layer are regarded as having pseudospin up and states in the bottom layer are 
regarded as having pseudospin down, the coherence factors for a pseudospin with orientation 
specified by polar and azimuthal angles $\theta$ and $\phi$ are given by the familiar spin-1/2 
coherent state expressions: $u_k = \cos(\theta/2)$ and $v_k = \sin(\theta/2) \exp(i \phi)$. Why, 
then, have we been describing them as excitonic condensates, which are expected to have superfluid
properties? The answer to this question appears to be\cite{bonsager} that ferromagnets can in 
principle exhibit superfluid behaviors similar to those of exitonic condensates, provided that they have 
nearly perfect easy-plane magnetic anisotropy. This kind of spintronic effect, in which the order 
parameter field is driven to a metastable configuration, is qualitatively different from giant 
magnetoresistance, current induced magnetization reversal, or any of the other interesting 
effects\cite{buhrman} that have been studied in ferromagnetic metals over the past decade. 
Advances in metal spintronics do however bring us closer to being able to realize the conditions 
required for their observation. In this section we switch from 
excitonic to magnetic language in briefly discussing one collective transport anomaly that has 
been seen in phase coherent bilayer quantum Hall systems. Our intention in doing so 
is to suggest that the transport anomalies very similar to those recently observed in quantum 
Hall bilayers, could also occur in thin film ferromagnets. Similar anomalies should also occur 
in zero field excitonic condensates when phase coherence is eventually achieved there; something 
that we await with confidence. 

We are interested in modeling two-terminal $I$-$V$ measurements in which a voltage difference is 
applied between contacts that are on opposite ends of the bilayer system and make contact only 
with top and bottom layers respectively. What was discovered experimentally\cite{jpeexpt} is 
that a large zero-bias peak emerges in the two-terminal conductance when interlayer 
phase coherence is established. In this paper, we discuss a purely one-dimensional model that, 
because of the importance of edge states in the quantum Hall regime, is not adequate to fully 
describe the experimental situation; work on a fully two-dimensional models that account for 
${\bf E}\cross{\bf B}$ drift near the edge of the sample is currently in progress. Despite its 
relative simplicity some ideas 
that we believe to be key emerge from the analysis described below. We describe collective 
transport\cite{jpeexpt} using equations of motion for the order parameter field that can be 
derived microscopically\cite{joglekardcj}, at least in the linear response regime. These 
equations are basically the Landau-Liftshitz equations that we expect for systems with magnetic 
order, but some features are special in the quantum Hall regime. We describe the order 
parameter field at a given position and time in terms of two variables $\phi$, the azimuthal 
angle of the pseudospin field, and its direction cosine along the polar direction $\Omega_z$. 
More physically, $\phi$ describes the relative phase between condensed electrons in the two 
layers while $\Omega_z$ describes the degree of charge transfer between the layers. In the 
excitonic condensate language $\phi$ is the phase of the electron-hole Cooper pair field. In the 
present analysis we assume that $\Omega_z$ is always small, {\it i.e.} that the fraction of the 
local charge density transferred from one layer to another is always small, but we allow $\phi$ 
to vary arbitrarily.

We have derived\cite{joglekardcj,tobepub} the following equations of motion for the order 
parameter field, 
\beq
             \frac{\partial\varphi}{\partial t} = -
              \frac{4\pi l^2\beta}{\hbar} \Omega_z   -
              \frac{\alpha_\varphi}{\hbar}
              \left[\frac{\Delta_t}{2}\sin\pt - 2 \pi l^2\rho_s\nabla^2\pt\right]
 \label{eqbasic}
\enq
\beq
              \frac{\partial\Omega_z}{\partial t} =
              \frac{1}{\hbar}
              \left[\frac{\Delta_t}{2}\sin\pt - 2 \pi l^2\rho_s\nabla^2\pt\right] 
            + \frac{8\pi^2 l^4\beta\sigma_z}{e^2 M_0}\nabla^2\Omega_z 
 \label{eqbasic2}
\enq 
where\cite{dlreview} $\rho_s$ is the pseudospin stiffness, $\beta$ is a measure of 
the capacitive energy cost of tilting the pseudospin out if its $x$-$y$ easy plane, $\Delta_t$ is 
the interlayer tunneling amplitude, $M_0$ is the order parameter amplitude, $\alpha_\varphi$ is a 
dimensionless Gilbert damping parameter for the pseudospin field, and $\sigma_z$ is a transverse 
pseudospin conductivity. In these equations $4\pi\ell^2\beta\Omega_z$ is the electrochemical 
potential difference between top and bottom layers, the $\hat z$ direction component of the 
pseudospin effective magnetic field.  The first term on the right hand side of Eq.~\ref{eqbasic} 
describes precession of the collective pseudospin order parameter field around the $z$ direction, 
while the second term describes damping processes which allow the $x$-$y$ component of the 
pseudospin to align with the $x$-$y$ component of the effective pseudospin field. The damping 
term is non-zero when this alignment is imperfect. Since inter-layer tunneling favors equal 
phases in the two-layers, it contributes a pseudospin-effective field in the $\hat x$ direction 
to the Hamiltonian; $\Delta_t \sin (\varphi)/2$ is the component of the pseudospin field which is 
out of alignment with a pseudospin that has azimuthal orientation $\varphi$. The second term 
arises from the microscopic 
exchange fields which produce pseudospin effective fields\cite{abolfath} that are oriented along 
a direction obtained by averaging over spatially nearby pseudospin orientations. When the 
pseudospin orientation varies spatially, this average will not be locally aligned with the 
pseudospin direction. 

The second equation describes the precession of the $\hat z$ component of 
the pseudospin field around the transverse $x$-$y$ plane field, and damping terms that allow 
this component of the pseudospin field to decay toward alignment with the effective magnetic 
field. An interesting aspect of the physics of these bilayer quantum Hall ferromagnets is that 
the Gilbert damping coefficient in this equation vanishes\cite{joglekardcj}. 
Since decay of the pseudospin current cannot occur locally, we include in our analysis a 
quasiparticle transport term, which allows the $\hat z$ component of the pseudospin field to 
decay when microscopic pseudospin-polarized currents have finite divergence. The three terms on 
the right hand side of Eq.~\ref{eqbasic2} can be identified as representing collective 
inter-layer tunneling, two-dimensional supercurrent, and quasiparticle current contributions to 
$\partial\Omega_z/\partial t$.

We believe that these Landau-Liftshitz equations describe the essence of order parameter dynamics 
in quantum Hall bilayer coherent states. In order to build a theory of two-terminal inter-layer 
transport $I$-$V$ curves, we must prescribe boundary conditions at the edge of the system. We 
focus here on balanced bilayers for which charge (sum of current in the two layers) and pseudospin 
polarized (difference of current in the two layers) currents can be considered separately. In the 
inter-layer tunneling geometry experiment we focus on the pseudospin-current which changes sign 
(in the top layer and out the bottom) across the sample. Since the two-terminal voltage is 
measured between a top layer contact on one side of the sample and a bottom layer contact on the 
other side of the sample, its pseudospin-contribution will be the sum of chemical potential 
differences between top and bottom layers on the two ends of the sample, which will identical by 
symmetry.   
\beq
 \left.\mu_z\right|_{-L/2} = \frac{eV_{PS}}{2}
 \label{eqbccpleft}
\enq
\beq
 \label{eqbccpright}
 \left.\mu_z\right|_{L/2} = \frac{eV_{PS}}{2}
\enq
where $\mu_z = 4 \pi \ell^2 \beta \Omega_z$, $V_{PS}$ is the pseudospin contribution to the 
measured voltage difference, and $L$ is the linear size of the sample. The total measured voltage 
also has a contribution from the charge current channel, but this will be much smaller; in the 
following we regard $V_{PS}$ as the total two-terminal voltage $V$. 

The large zero-bias peak in the two-terminal conductance can be understood in terms of stationary 
solutions of these equations for collective dynamics. We find that the time-independent 
pseudospin fields satisfy 
\beq
 \frac{\diff^2\Omega_z}{\diff x^2} -\frac{1}{L_z^2}\Omega_z = 0
 \label{eqstmz}
\enq
\beq
 \frac{\diff^2\pt}{\diff x^2} - \frac{1}{\lambda_j^2}\sin\pt =
 \frac{2\beta}{\alpha_\varphi\rho_s}\Omega_z
 \label{eqstph}
\enq
where we have introduced the two length scales defined by these equations. $\lambda_j$ is a 
domain wall width in the magnetic language and a {\it Josephson length} in the excitonic 
superfluid language. It is defined by
$$
 \lambda_j = l\sqrt{|\frac{4\pi\rho_s}{\Delta_t}}.
$$
The length $L_z$ is discussed below. These equations can be solved using the boundary condition 
that $d \varphi/ dx$ vanish that both ends of the sample, {\it i.e.} that no supercurrent flow 
into or out of the sample. This leads to 
\beq
\Omega_z = \frac{eV}{8\pi l^2\beta\cosh\left(\frac{L}{2L_z}\right)}\cosh\left(\frac{x}{L_z}\right) 
 \label{eqstmzsol}
\enq
where the length $L_z$ is defined by:
$$
 L_z = l \sqrt{\frac{2\pi\hbar\sigma_z\alpha_\varphi}{e^2 M_0\left(\alpha_z +
\alpha_z\alpha_\varphi\right)}}
$$
$L_z$ is the length scale for relaxation of the pseudospin polarization of the quasiparticle 
current injected from the contacts. In the limit $L_z\ll\lambda_j$, the stationary solution to 
the damped Landau-Liftshitz-Gilbert equations reflects a process in which the injected 
quasiparticle pseudospin current is converted into a supercurrent, which then decays by 
collective interlayer tunneling. These solutions exist only up to a critical value of the 
source-drain voltage $V$, explaining the sharp peak in the two-terminal conductivity near zero 
bias. $I(V)$ can be evaluated from the following expression for the purely quasiparticle currents 
entering and exiting the sample:  
$$
 j = \frac{4\pi l^2\beta\sigma_z}{e}\left|\left.\frac{\partial\Omega_z}{\partial x}\right|_
{\pm L/2}\right.
$$
It follows that the two-terminal resistance of this theory is proportional to the product of the 
quasiparticle resistivity and the length $L_z$ over which the quasiparticle current is present. 
Beyond the critical voltage only time-dependent solutions to these equations exist and the time 
averaged current decreases. 


\section{Summary} 
Bilayer quantum Hall systems near total filling factor $\nu =1$ have broken symmetry ground 
states with spontaneous interlayer phase coherence between two-dimensional subbands that are 
localized in different layers. Spontaneous coherence between different bands is the property that 
defines excitonic Bose condensation, leading for example to superfluid behavior for currents 
flowing in opposite directions in the two bands. The recent observation of anomalous collective 
transport effects in the quantum Hall regime when bilayer coherence is present therefore 
represents the experimental discovery of the long-sought excitonic Bose-Einstein condensate. In 
this paper we have briefly discussed the possibility of similar superfluid collective transport 
effects in thin-film ferromagnets with nearly-perfect easy-plane anisotropy and efforts to 
develop a complete picture of two-terminal bilayer quantum Hall transport measurements in which 
current is injected into one of the electron layers and removed from the other.


\section*{Acknowledgments}
We are grateful for helpful interactions with Ramin Abolfath, Steve Girvin, Tomas Jungwirth, and 
Ady Stern. The work was supported by the Welch Foundation and by the National Science Foundation 
under grant DMR-0115947.


\end{document}